\documentclass[preprint,showkeys]{revtex4} 
\usepackage{amsmath}
\usepackage{latexsym}
\usepackage{amssymb}
\usepackage{graphics,epstopdf}
\usepackage{amsmath,amssymb,amsthm}
\usepackage{graphicx}
\usepackage[colorlinks=true, citecolor=blue, urlcolor=blue ]{hyperref}
\usepackage{cleveref}
\newtheorem{theorem}{Theorem}

\begin{document}
	\title{Information geometry of entangled states induced by noncommutative deformation of phase space}
	\author{Shilpa Nandi}
	\email{nandishilpa801325@gmail.com}
	\affiliation{Department of Physics, Brahmananda Keshab Chandra College, 111/2 B. T. Road, Kolkata, India-700108}
	\author{Pinaki Patra}
	\thanks{Corresponding author}
	\email{monk.ju@gmail.com}
	\affiliation{Department of Physics, Brahmananda Keshab Chandra College, 111/2 B. T. Road, Kolkata, India-700108}
	
	\keywords{Information geometry; Gaussian entanglement; Noncommutative space; Fisher-Rao metric}
	
	\date{\today}

	\begin{abstract}
		In this paper, we revisit the notion of quantum entanglement induced by the deformation of phase-space through noncommutative space (NC) parameters. The geometric structure of the state space for Gaussian states in NC-space is illustrated through information geometry approach. We parametrize the phase-space distributions by their covariances and utilize the Fisher-Rao metric to construct the statistical manifold associated with quantum states. We describe the notion of the Robertson-Scr\"{o}dinger uncertainty principle (RSUP) and positive partial transpose (PPT) conditions for allowed quantum states and separable states, respectively, for NC-space. RSUP and PPT provide the restrictions on all allowed states and separable states, respectively. This enables us to estimate the relative volumes of set of separable states and entangled states.   Numerical estimations are provided for a toy model of a bipartite Gaussian state. We restrict our study to such bipartite Gaussian states, for which the entanglement is induced by the noncommutative phase-space parameters.
	\end{abstract}
	
	\maketitle
	
	\section{Introduction}
	More than a century ago, Einstein put forth the current theory of gravity in its final form: General Relativity (GR), which has passed every observational and experimental test it has ever faced \cite{gr1}. On the other hand, we have not witnessed any violation of predictions of another century-old theory, namely Quantum mechanics (QM). GR and QM are two cornerstones of our modern understanding of nature.  Despite their success in their region of validity, QM and GR are not consistent with each other.  Although, there are several beautiful mathematically consistent proposals for quantum theory of gravity \cite{qg1,qg2}, none of them have been successfully tested experimentally, to date. One possible reason is the huge gap between the required energy scale ($(10^{19})$ GeV)  and the currently achievable energy scales ($14$ TeV for LHC \cite{lhc}) to test predictions of quantum gravity (QG) through collider physics \cite{qg3}. There is a consensus among most of the theories of QG that the space-time structure is deformed in such a manner that the usual notion of commutative space is ceased at the energy scales in which it is predicted to show departures from GR \cite{qg4,qg5}. In particular,  the fundamental concept of space-time is mostly compatible with quantum theory in noncommutative (NC) space \cite{qg6,qg7}. In other words, at very high energies, a common expectation is that space-time may not retain its smooth continuous structure at very short distances \cite{qg8,ncs1}.
	Moyal deformation of ordinary space-time is an example of a specific algebraic realization of this expectation \cite{ncs2}. Not only in high-energy regimes but also in low-energy situations, such as in the quantum Hall effect,  the presence of a magnetic field makes the guiding center coordinates of the electron noncommutative \cite{ncs3}. It is thus crucial to explore the aspects of physical phenomena under the NC deformation of phase space.\\
	In the present paper,  we wish to address a novel aspect of QM in NC space, namely quantum entanglement \cite{entangle1}.  Quantum entanglement is the key ingredient for the storage and distribution of quantum information in the quantum world \cite{entangle2}. The nontrivial consequences of entanglement on quantum ontology is an enriched and age-old well-known issue. For instance, it was noticed nine decades ago within the context of position-momentum continuous-variable systems by  Einstein,  Podolsky, and  Rosen \cite{entangle3}. In recent times there has been a rapid development of the theory of information-processing applications, such as quantum teleportation \cite{entangle4}, dense coding \cite{entangle5}, and cryptography \cite{entangle6},
	which have been extended for continuous variable systems. One of the fundamental questions concerning these subjects is to estimate how many entangled states exist among all quantum states. In particular, one might quest for the answer to the philosophical question \textquotedblleft Is the world more classical or more quantum\textquotedblright?  From a practical point of view, entangled states are the key ingredients for any quantum protocols. Therefore, obtaining a relative measure of entangled states overall quantum states is important in its own right.  \\
	Here, we explore how the geometric properties of entanglement and information are affected by the noncommutativity of the phase space. Through information geometry, we estimate a relative measure of entangled states in NC space. 
	In statistics, information geometry analyses the space of parametric probability distributions by introducing notions of a metric.  Extension of  concepts of information geometry in quantum mechanical systems has been found to have fruitful applications in the foundations of quantum theory \cite{infogeo1,infogeo2}, metrology \cite{infogeo3}, quantum thermodynamics \cite{infogeo4}, phase transitions \cite{infogeo5}, random state preparation \cite{infogeo6}
	and quantum speed limits\cite{infogeo7}. The central idea of the formalism of quantum information geometry is to construct a Riemannian metric on the space of parameterized quantum states \cite{infogeo8}. This gives rise to a notion of statistical distance between different states, where length is assigned based upon their degree of distinguishability \cite{infogeo9}. The set of quantum states is a convex set. The pure states are its extreme points \cite{sep1}. Fubini-Study metric provides a straightforward way for the geometrical settings of pure states in continuous variable systems \cite{fubini1,fubini2}. Moreover, for pure states, the Fubini-Study metric is equivalent to Fisher-Rao metric \cite{fubini2}. For mixed states, there is a lack of consensus in the formulation of geometrical structure for continuous variable systems in general \cite{geo1}. However, for Gaussian states, a geometrical notion to determine the relative measure of separable and entangled states is explicitly provided in \cite{geo2}. Gaussian states are those, for which the characteristic function is a Gaussian function of the phase space coordinates.  For a comprehensive study on Gaussian states, one can see \cite{gaussian1}.  The simplicity of a Gaussian state is that it can be completely determined by its covariance matrix and first-moment vector. First-moment vector can be set to zero for all practical purposes with the help of coordinate shifting. In that case, one can associate Fisher-Rao information matrix $g(\theta)$ with elements $g_{\mu\nu}= \frac{1}{2}\mbox{Tr}[V^{-1}(\partial_\mu V) V^{-1}(\partial_\nu V)]$, with the covariance matrix $V$ of the Gaussian state $\hat{\rho}$. Here $\partial_\mu $ stands for the partial derivative with respect to the nonzero independent entities $\theta_\mu \in \Theta$ of $V$. The parameter space $\Theta$ is constrained by the fact that there exists measurement unsharpness among canonically conjugate observables. In other words, $\Theta$ is restricted by the Robertson-Schr\"{o}dinger uncertainty principle $V+\frac{i}{2}J\ge 0$, where the elements $J_{\alpha\beta}$ of the $2n\times 2n$ symplectic matrix $J$ are given by the commutation relations $[\hat{X}_\alpha,\hat{X}_\beta]$ of the phase-space operators $X=(\hat{x}_1,...\hat{x}_n,\hat{p}_1,...\hat{p}_n)$. We have considered the Planck constant $\hbar=1$, which will be followed throughout the present paper unless otherwise specified. One can endow the parameter space $\Theta$ with a Riemannian metric, namely the Fisher-Rao metric, given by  $G(\theta)=\sum_{\mu,\nu}g_{\mu\nu}d\theta^\mu \otimes d \theta^\nu$, which induced a Riemannian
	manifold $\mathcal{M}=(\Theta, G(\theta))$, on which one can introduce a volume measure $\Gamma(V)= \int_{\Theta}d\theta \sqrt{\mbox{det}(g(\theta))}$.\\
	Separable states (not entangled) are the states of a composite system that can be written as convex combinations of subsystem states \cite{sep2}. For discrete quantum systems, the positive partial transpose (PPT) criterion provides a necessary (for Gaussian states it is sufficient) condition to determine the separability of bipartite states \cite{sep3,sep4}. The idea of PPT was extended for continuous variable systems by implementing partial transposition operation as a mirror reflection in the Wigner phase-space \cite{sepcont1,sepcont2}.  Formalism of determination of entangled states through generalized PPT criteria in NC-phase space is introduced in \cite{ncseparability1} and further studied in literature \cite{ncseparability2,ncseparability3}.  PPT condition for separable states provides constraints on the allowed parameter space of the covariance matrix. Evaluating the volumes of states under the constraints, one can then estimate a measure of separable states.  The relevance of considering Gaussian states is twofold: first,	Gaussian states are the most commonly experimentally used continuous-variable states. Second, Gaussian quantum states are represented in the phase space picture of quantum mechanics as a proper probability density function.\\
	The organization of the present paper is as follows. First, we describe the entanglement induced by the noncommutative parameters for bipartite Gaussian states. Then we discuss the geometry of Gaussian states through Fisher-Rao metric. Then we illustrate the effects of deformation of phase-space on the geometry of Gaussian states. We use a toy model for explicit numerical results. Finally, we discuss the aspects of our findings in conclusions.
	%%%%%%%%%%%%%%%%%%%%%%%%%%%%%%%%%%%%%%%%%%%%%%%%%%%%%%%%%%%%%%%%%%%%%%%%%%%%
	%%%%%%%%%%%%%%%%%%%%%%%%%%%%%%%%%%%%%%%%%%%%%%%%%%%%%%%%%%%%%%%%%%%%%%%%%%%%%%
	\section{Entanglement induced by noncommutative parameters}
Let us consider that a bipartite quantum system with $2n=2n_A+2n_B$ dimensional phase space is described by two subsystems Alice (A) and Bob (B), who have two quantum systems of $2n_A$ and $2n_B$ dimensional phase space, respectively.
Let us consider that A and  B share a Gaussian state $\rho(z)$ in noncommutative (NC) space, in which generalized coordinates satisfy the commutation relations
\begin{equation}
	[\hat{z}_j,\hat{z}_l]=i\Omega_{jl},\;\;\; j,l=1,2,...,2n,
\end{equation}
through the deformed symplectic matrix
\begin{equation}
	\Omega=[\Omega_{ij}]\equiv \mbox{Diag}[\Omega^A,\Omega^B].
\end{equation}
 $\Omega^K$ is a real skew-symmetric nonsingular $2n_k\times 2n_k$ matrix of the form
\begin{eqnarray}
	\Omega^K=\left(\begin{array}{cc}
		\Theta^K & \mathbb{I}^K \\
		-\mathbb{I}^K & \Upsilon^K
	\end{array}\right),\; K=A,B.
\end{eqnarray}
 $\Theta^K=[\theta_{ij}^K]_{i,j=1}^{n_K}$ and $\Upsilon^K=[\eta_{ij}^K]_{i,j=1}^{n_K}$ encode measure of the noncommutativity of position ($\hat{x}_i^K$) - position and momentum  ($\hat{p}_i^K$) - momentum sector, respectively. $\mathbb{I}^K$ stands for the $n_k\times n_K$ identity matrix. 
 Here we have used shorthand notation $\hat{z}=(\hat{z}^A,\hat{z}^B)$, with $\hat{z}^K=(\hat{x}_1^K,...,\hat{x}_{n_K}^K, \hat{p}_1^K,...,\hat{p}_{n_K}^K)$ corresponds to the co-ordinates of $K=A,B$. \\
 The NC-structure of phase-space may be formulated in terms of equivalent usual commutative variables $\hat{\xi}=(\hat{\xi}^A,\hat{\xi}^B)$, through a linear Darboux transformation (DT)
\begin{equation}\label{darbouxtransformation}
	\hat{z}=S\hat{\xi},
\end{equation}
where $\hat{\xi}=(\hat{\xi}^A,\hat{\xi}^B)$, with
$
\hat{\xi}^K=(\hat{q}_1^K,..\hat{q}_{n_K}^K, \hat{k}_{1}^K,..\hat{k}_{n_K}^K),\; K=A,B.
$
Commutative space co-ordinates ($\hat{q}_j^K, \;K=A,B$) and momentum ($\hat{k}_{j}^K,\; K=A,B$) satisfy the usual canonical commutation relations
\begin{equation}
	[\hat{\xi}_j,\hat{\xi}_l]=iJ_{jl},\; j,l=1,2,...,2n,
\end{equation}
where $J_{ij}$ is the $ij^{th}$ element of the symplectic matrix
$J=\mbox{Diag}[J^A,J^B]\in Sp(2n,\mathbb{R})$, with
\begin{eqnarray}
	J^K= \left(\begin{array}{cc}
0 & \mathbb{I}^K \\
-\mathbb{I}^K & 0
	\end{array}
\right)\in Sp(2n_K,\mathbb{R}),K=A,B.
\end{eqnarray}
The linear transformation
\begin{equation}
	S=\mbox{Diag}[S^A,S^B] \in Gl(2n,\mathbb{R})
\end{equation}
is given through $\Omega^K$ and $J^K$ as
\begin{eqnarray}
	\Omega^K &=&S^K J^K (S^K)^T,\; K=A,B.\\
	\implies &&	\Omega=S J S^T. \label{OmegaJconnection}
\end{eqnarray}
The shared bipartite state $\rho(z)$ in NC space is equivalent to the commutative space  bipartite  state $\tilde{\rho}(\hat{\xi})$, which is obtained through the Darboux transformation ~\eqref{darbouxtransformation} as
\begin{equation}
\tilde{\rho}(\hat{\xi})=\rho(S\hat{\xi}),
\end{equation}
which is associated with the Wigner function
\begin{eqnarray}\label{commutativewigner}
W_{\tilde{\rho}}(\xi) = \frac{1}{(2\pi)^n}\int_{\mathbb{R}^{n_{A}}} dy^A \int_{\mathbb{R}^{n_{B}}} dy^B    \langle q^A+\frac{y^A}{2}, q^B+\frac{y^B}{2} \vert \tilde{\rho} \vert q^A-\frac{y^A}{2},q^B-\frac{y^B}{2} \rangle \nonumber \\
 e^{-i(y^A.k^A+y^B.k^B)}.
\end{eqnarray}
Corresponding NC space Wigner function reads
\begin{equation}\label{NCwigner}
	W_{\rho}(z)= \frac{1}{\mbox{det}(\Omega)}W_{\tilde{\rho}}(S^{-1}z).
\end{equation}
The covariance matrix $\Sigma$ corresponding to the NC space Wigner distribution $W_\rho$ is related to the covariance matrix $\tilde{\Sigma}$ of commutative space distribution $W_{\tilde{\rho}}$ as follows.
	\begin{equation}\label{sigmasigmatildeconnection}
		\Sigma=S\tilde{\Sigma}S^T.
	\end{equation} 
Every bonafide covariance matrix in usual commutative space  satisfy the Robertson-Schr\"{o}dinger uncertainty principle (RSUP)
\begin{equation}\label{RSUPC}
	\tilde{\Sigma}+\frac{i}{2}J\ge 0.
\end{equation}
Using ~\eqref{OmegaJconnection}, ~\eqref{sigmasigmatildeconnection} and ~\eqref{RSUPC}, we get the NC-space RSUP
\begin{equation}\label{NCRSUP}
	\Sigma+ \frac{i}{2}\Omega \ge 0.
\end{equation}
One of the equivalent statements of the RSUP ~\eqref{RSUPC} is that the smallest eigenvalue $\tilde{\nu}_{-}$ of $2i\tilde{\Sigma}$ greater than one, i.e., the smallest symplectic eigenvalue of $\tilde{\Sigma}$ satisfies $\tilde{\nu}_{-}\ge 1$. At the same token, if we define the NC symplectic eigenvalues of $\Sigma$ as the eigenvalues of $2i\Omega^{-1}\Sigma$, then NC RSUP~\eqref{NCRSUP} can be stated in terms of the smallest eigenvalue $\nu_{-}$ of $2i\Omega^{-1}\Sigma$ as $\nu_{-}\ge 1$.
One of the immediate implications of the relations ~\eqref{OmegaJconnection} and ~\eqref{sigmasigmatildeconnection} is the following.
\begin{theorem}[Symplectic spectrum]\label{sympecticspectrum}
	NC symplectic spectrum of $\Sigma$ and the symplectic spectrum of $\tilde{\Sigma}$ are identical.
\end{theorem}
Theorem~\ref{sympecticspectrum} can be envisaged directly from the characteristic polynomial of the symplectic spectrum of $\tilde{\Sigma}$, i.e., the ordinary spectrum of $iJ^{-1}\tilde{\Sigma}$. In particular, we can write the characteristic polynomial as
\begin{eqnarray}
	\mbox{Det}(2iJ^{-1}\tilde{\Sigma}-\lambda\mathbb{I}) &=& \mbox{Det}(S^{T})^{-1}	\mbox{Det}(2iJ^{-1}\tilde{\Sigma}-\lambda\mathbb{I})\mbox{Det}(S^{T}) \nonumber \\
	&=& \mbox{Det}(2i (S^T)^{-1}J^{-1}S^{-1}\Sigma (S^T)^{-1}S^T-\lambda (S^T)^{-1}S^T) \nonumber \\
	&=&\mbox{Det}(2i\Omega^{-1}\Sigma -\lambda\mathbb{I}),
\end{eqnarray}
which means the characteristic polynomials of $2iJ^{-1}\tilde{\Sigma}$ and $2i\Omega^{-1}\Sigma$ coincides.
If the shared state $\rho$ is separable, then it can be expressed as
\begin{equation}
	\rho=\sum_{j=1}^{\infty} \lambda_j \rho_j^A\otimes\rho_j^B,\; 0\le \lambda_j\le 1, \; \sum_{j=1}^{\infty}\lambda_j=1.
\end{equation}
Here $\rho_j^A$ and $\rho_j^B$ corresponds to density matrix involves only coordinates of Alice ($\hat{z}^A$) and Bob  ($\hat{z}^B$), respectively. The associated phase space distribution may be expressed as 
\begin{equation}
	W_\rho(z)= \sum_{j=1}^{\infty} \lambda_j W_{\rho_j^A}(z^A)W_{\rho_j^B}(z^B),
\end{equation}
which corresponds to the following distribution function in commutative space.
\begin{equation}
		W_{\tilde{\rho}}(\xi)= \sum_{j=1}^{\infty} \lambda_j W_{\tilde{\rho}_j^A}(\xi^A)W_{\tilde{\rho}_j^B}(\xi^B).
	\end{equation}
For a general class of systems, there is no sufficient condition to verify the separability condition, even for a bipartite system. However, for Gaussian states, one can use the positive partial transpose (PPT) criterion to identify separable states. 	Let a $2n\times 2n $ matrix $\Lambda=\mbox{Diag}(\mathbb{I}^A,\Lambda^B)$, with $\Lambda^B=\mbox{Diag}[\mathbb{I},-\mathbb{I}]$ denotes the partial transposition operation (in this case $\xi \mapsto \Lambda\xi$ is the mirror reflection of Bob's momentum) concerning the party B. PPT criterion for separability states followings.
\begin{theorem}[PPT Condition]
	Under a PPT operation the Wigner distribution $W_{\tilde{\rho}}(\xi)$ corresponding to a separable state transforms to an equally admissible Wigner distribution. 
\end{theorem}
That means after PPT transformation on a separable state
\begin{equation}\label{Wtranformlambda}
W_{\tilde{\rho}}(\xi) \mapsto W_{\tilde{\rho}'}(\xi)=W_{\tilde{\rho}}(\Lambda\xi),
\end{equation}
has a bonafide Wigner distribution on phase space. In other words, for the separable state $\tilde{\rho}$, the covariance matrix transforms to a bonafide covariance matrix after a PPT operation. Hence, the  covariance matrix $\tilde{\Sigma}'$ corresponding to $W_{\tilde{\rho}'}(\xi)$, satisfy the RSUP 
\begin{equation}
	\tilde{\Sigma}'+\frac{i}{2}J\ge 0.
\end{equation}
Using ~\eqref{Wtranformlambda}, ~\eqref{NCwigner} and ~\eqref{sigmasigmatildeconnection}, one can show that after PPT transformation the RSUP for NC-space reads as
\begin{equation}\label{NCRSUPPPT}
\Sigma'+\frac{i}{2}\Omega\ge 0,\; \mbox{where}\; \Sigma'=S\tilde{\Sigma}'S^T.
\end{equation}
Moreover, we have 
\begin{equation}
	\Sigma'=D\Sigma D^T, \;\mbox{with}\; D=D^{-1}=S\Lambda S=\mbox{Diag}[\mathbb{I}^A, S^B\Lambda^B (S^B)^{-1}],
\end{equation}
which simplifies the RSUP after PPT in NC space as
\begin{equation}
	\Sigma+\frac{i}{2}\Omega'\ge 0,\; \mbox{with}\; \Omega'=D^{-1}\Omega (D^T)^{-1}=\Omega'=\mbox{Diag}[\Omega^A,-\Omega^B].
\end{equation}
If $\nu'_{-}$ is the smallest eigenvalue of $\Sigma'$, then the statement ~\eqref{NCRSUPPPT} is equivalent to $\nu'_{-}\ge 1$.
Hence, we arrive at the following simple condition for a separable bipartite Gaussian system.
\begin{theorem}[Separability Condition for Gaussian system in NC space]
	If a real normalizable bipartite NC phase-space Gaussian distribution with covariance matrix $\Sigma$ is separable, then the following statements are equivalent.
	\begin{itemize}
		\item  Both the conditions $\Sigma+\frac{i}{2}\Omega\ge 0$ and  $\Sigma+\frac{i}{2}\Omega'\ge 0$ are satisfied. 
		\item $\nu_-\ge 1$ and $\nu'_- \ge 1$.
	\end{itemize}
\end{theorem}
\section{Fisher-Rao metric for Gaussian states in NC-space}
One can categorize the class of separable states and entangled states with the help of PPT criterion. This enables us to measure the set of entangled states out of all quantum states. To do so, let us first parametrize the  Gaussian probability density function $P(\xi)$,  by the nonzero  entities  of the covariance matrix as 
\begin{equation}
	\mathcal{S}:=\left\{ P(\xi)\equiv P(\xi;\theta)=\frac{e^{-\frac{1}{2}z^T \Sigma^{-1}(\theta)z}}{(2\pi)^n\sqrt{\mbox{det} \Sigma(\theta)}},\vert \theta \in \Theta\right\}.
\end{equation}
The parametrization $\theta=(\theta_1,....,\theta_m)$ is provided through elements $\Sigma_{\mu\nu}$ of $\Sigma$ as
\begin{equation}
	\theta_l=\Sigma_{\mu\nu},\; \mbox{with}\; l=\sum_{r=0}^{\mu-2} (2n-r)+\nu-\mu+1,\; 1\le l\le m.
\end{equation}
We note that $\theta \mapsto P(.;\theta)$ is injective and $\Theta \subseteq \mathbb{R}^m$ is obtained from specific constraint on $\Sigma(\theta)$. For instance, for a two-mode bipartite separable system, the constraint is provided by ~\eqref{NCRSUP}. In general, the parameter space $\Theta$ for quantum states in NC-space is 
\begin{equation}\label{thetaquantum}
	\Theta_{\mbox{quantum}}^{NC}:= \left\{\theta\in \mathbb{R}^m\vert \Sigma(\theta)+\frac{i}{2}\Omega\ge 0\right\},
\end{equation}
whereas for a bipartite separable state 
\begin{equation}\label{thetaseparable}
	\Theta_{\mbox{separable}}^{NC}:= \left\{\theta\in \mathbb{R}^m\vert \Sigma+\frac{i}{2}\Omega'\ge 0\right\}.
\end{equation}
If we assume that, a quantum state is entangled if it is not separable, then
\begin{equation}\label{thetaentangled}
	\Theta_{\mbox{entangled}}^{NC}:= \Theta_{\mbox{quantum}}^{NC}-\Theta_{\mbox{separable}}^{NC}.
\end{equation}  
The Fisher information matrix of $\mathcal{S}$ at $\theta\in \Theta$ is the $m\times m$ matrix $g(\theta)$ with elements
\begin{equation}\label{gmunudefn}
	g_{\mu\nu}(\theta):= \int_{\mathbb{R}^{2n}} P(\xi;\theta)\partial_\mu \ln P(\xi;\theta)\partial_\nu \ln P(\xi;\theta) d\xi,
\end{equation}
where $\partial_\mu$ stands for $\frac{\partial}{\partial\theta_\mu}$. 
In terms of the covariance matrix, ~\eqref{gmunudefn} can be rewritten as
\begin{equation}\label{gmunuV}
	g_{\mu\nu}= \frac{1}{2}\mbox{Tr}[\Sigma^{-1}(\partial_\mu \Sigma)\Sigma^{-1}(\partial_\nu \Sigma)].
\end{equation}
~\eqref{gmunuV} is computationally simpler than that of ~\eqref{gmunudefn}.
The positive semidefinite matrix $g(\theta)$ endows the parameter space $\Theta$ with a Riemannian metric (the Fisher-Rao metric) 
\begin{equation}
	G(\theta):= \sum_{\mu\nu} g_{\mu\nu}(\theta)d\theta^{\mu}\otimes d\theta^{\nu}.
\end{equation}
On the Riemannian manifold $\mathcal{M}:= (\Theta, G(\theta))$, associated with the class of Gaussian state parametrized by $\Theta$, we can define the volume of the physical states represented by $\Theta$ as
\begin{equation}\label{volumedefn}
	\Gamma_\mathcal{V}:=\int_\Theta v_G,\; \mbox{with}\; v_G:=\sqrt{\mbox{det} g(\theta)} d\theta_1\wedge.... \wedge d\theta_m.
\end{equation}
The volume measure ~\eqref{volumedefn} enables us to compare the measure of entangled state with that of separable one, quantitatively. However, one can note that $\mbox{det}g(\theta)$ diverges for some $\theta_l\in\Theta$. In particular, since $\mbox{det}g(\theta)=\tilde{F}(\Sigma(\theta))(\mbox{det}\Sigma(\theta))^{-2m}$, with $\tilde{F}(\Sigma(\theta))$ a non-rational function of $\theta_1,...,\theta_m$, $\mbox{det}g(\theta)$ diverges at zeroes of $\Sigma(\theta)$. To overcome the issue of possible divergence of the volume, we introduce a regularization function. We require that the regularization function be invariant under symplectic transformation $S\in Sp(2n,\mathbb{R})$. The following choice of regularizing function serves the required purpose.
\begin{equation}\label{regularizationfunction}
	\Upsilon(\Sigma) := e^{-\frac{1}{\kappa}\mbox{Tr}[adj(\Sigma)]} \log[1+(\mbox{det}\Sigma)^m].
\end{equation}
Here $adj(\Sigma)$ denotes adjunct of $\Sigma$. With the regularizing function ~\eqref{regularizationfunction} the volume of the set of Gaussian states represented by the parameter space $\Theta$ is thus defined by
\begin{equation}\label{volumeonconstraint}
	\Gamma_{\Upsilon}(\Sigma):= \int_\Theta \Upsilon(\Sigma) v_G.
\end{equation}
Now we observe that the equivalent Fisher-Rao metric in usual commutative space may be defined as
\begin{equation}\label{FRinc}
	\tilde{g}_{\mu\nu}=\frac{1}{2}\mbox{Tr}[\tilde{\Sigma}^{-1}(\tilde{\partial}_\mu \tilde{\Sigma})\tilde{\Sigma}^{-1}(\tilde{\partial}_\nu \tilde{\Sigma})],
\end{equation}
where $\tilde{\partial}_\mu := \frac{\partial}{\partial \tilde{\theta}_\mu}$, with $\tilde{\theta}_\mu$ being the nonzero elements of covariance matrix $\tilde{\sigma}$ in commutative space. Using ~\eqref{sigmasigmatildeconnection} in ~\eqref{FRinc}, we see that
\begin{equation}\label{gmunudefnnc}
	g_{\mu\nu}=\frac{1}{2}\mbox{Tr}[(S^T)^{-1}\tilde{\Sigma}^{-1}\frac{\partial\tilde{\Sigma}}{\partial\theta_\mu}\tilde{\Sigma}^{-1}\frac{\partial\tilde{\Sigma}}{\partial\theta_\nu}S^T]= \frac{1}{2}\mbox{Tr}[\tilde{\Sigma}^{-1}\frac{\partial\tilde{\Sigma}}{\partial\theta_\mu}\tilde{\Sigma}^{-1}\frac{\partial\tilde{\Sigma}}{\partial\theta_\nu}]=\tilde{g}_{\mu \nu}.
\end{equation}
In other words, the geometry of entangled states remains unchanged under the Darboux transformation. On the other hand, we observe the scaling property of the covariance matrix on the Fisher-Rao metric as follows. If the covariance matrix $\Sigma$ is a multiple of some another covariance matrix $\Sigma^0$ as follows
\begin{equation}
	\Sigma= \frac{1}{2} b \Sigma^0 ,
\end{equation}
then $g_{\mu\nu}$ is given in terms of the metric $g^0_{\mu\nu}$ corresponding to $\Sigma^0$ as
\begin{equation}\label{gmunub}
g_{\mu\nu}=g^0_{\mu\nu}+ b_{\mu\nu},
\end{equation}
where
\begin{eqnarray}
	g^0_{\mu\nu} &=& \frac{1}{2}\mbox{Tr}[(\Sigma^0)^{-1}(\partial_\mu \Sigma^0) (\Sigma^0)^{-1}(\partial_\nu \Sigma^0)], \label{gmunu0}\\
	b_{\mu\nu} &=& n(\partial_\mu \ln b)( \partial_\nu \ln b)+\frac{1}{2} [(\partial_\mu \ln b)\mbox{Tr}(V^{-1} \partial_\nu V)+ (\partial_\nu \ln b)\mbox{Tr}(V^{-1} \partial_\mu V)]. \label{bmunu0}
\end{eqnarray}
Let us summarize the algorithm of the scheme to evaluate the relative measure of states.
\begin{itemize}
	\item Evaluate the covariance matrix $\Sigma$ corresponding to the bipartite Gaussian state $\hat{\rho}(z)$.
	\item Parametrize ($\theta_\mu$) the independent elements of $\Sigma$.
	\item Compute Fisher-Rao metric $g(\theta)$ through ~\eqref{gmunuV}.
	\item Utilize ~\eqref{regularizationfunction} to get a regularization factor $\Upsilon(\Sigma)$.
	\item Constraints ~\eqref{thetaquantum} and ~\eqref{thetaseparable} provide the required parameter space for allowed quantum states and separable states, respectively.  
	\item Now the integrations ~\eqref{volumeonconstraint} on the region provided by the parameter spaces  ~\eqref{thetaquantum} and ~\eqref{thetaseparable} provide required volume for the state space corresponding to all quantum and separable states, respectively. Entangled states volume is the difference between quantum and separable states.
\end{itemize}
\section{Illustrative toy model}
\begin{figure}
	\includegraphics[width=8cm]{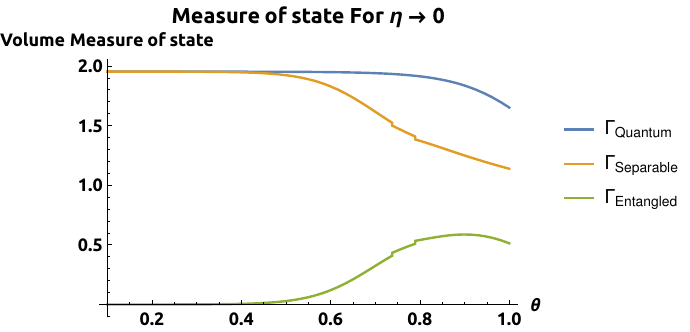}
	\caption{\textcolor{blue}{Measure of allowed states with respect to $\theta \in (0.1,1)$. The measure of the entangled state increases with $\theta$, as desired since the entanglement is induced with the deformation of phase-space. Here, we take the scale factor $\kappa \to 4$.}}
	\label{volrelative}
\end{figure}

For an illustrative example, let us consider an eight dimensional noncommutative phase space, for which the noncommutative structure of phase space is encoded in $\Omega=\mbox{Diag}[\Omega^A,\Omega^B]$, through 
\begin{eqnarray}
	\left[\hat{x}^A_r,\hat{x}^B_s \right] = i \theta  \delta_{AB} \epsilon_{rs}, \; r,s=1,2 . \\
	\left[\hat{p}^A_r,\hat{p}^B_s\right] = i \eta  \delta_{AB} \epsilon_{rs}, \; r,s=1,2 .\\
	\left[\hat{x}^A_r,\hat{p}^B_s\right] = i   \delta_{AB} \delta_{rs}, \; r,s=1,2 .
\end{eqnarray} 
Here we have used the notation $\delta_{ab}$ for Kronecker delta, which is equal to one for $a=b$, and zero otherwise. Total antisymmetric  Levi-Civita symbol $\epsilon_{ij}$ is set with the convention $\epsilon_{12}=-\epsilon_{21}=1$. Here we have considered $n_A=n_B=2$.
In particular, we consider
\begin{eqnarray}
	\Omega^A=\Omega^B= \left(\begin{array}{cc}
i\theta\sigma_y & \mathbb{I}_2 \\
-\mathbb{I}_2 & i\eta\sigma_y
	\end{array}\right),
\end{eqnarray}
where $\mathbb{I}_n$ is the $n\times n$ identity matrix and we take the representation of Pauli matrices as
\begin{eqnarray}
	\sigma_x = \left(\begin{array}{cc}
		0 & 1 \\
		1 &0
		\end{array}\right),\; 
		\sigma_y = \left(\begin{array}{cc}
		0 & -i \\
		i &0
	\end{array}\right),\;
	\sigma_z = \left(\begin{array}{cc}
	1 & 0 \\
	0 & -1
\end{array}\right).
\end{eqnarray}
The Darboux transformation (Bopp shift) corresponds to $S=\mbox{Diag}[S^A,S^B]$, with
\begin{eqnarray}
	S^A=S^B= \left(\begin{array}{cc}
\lambda \mathbb{I}_2 & -\frac{i\theta}{2\lambda}\sigma_y \\
\frac{\eta}{2\mu}\sigma_x & \mu\mathbb{I}_2
	\end{array}\right),\; \mbox{where}\; \lambda\mu =\frac{1}{2}(1+ \sqrt{1-\eta\theta});\; \theta\eta <1.
	\end{eqnarray}
One can verify that the transformation is invertible. Let us consider that,  Alice and Bob share the following Gaussian state on the noncommutative space.
\begin{equation}\label{sharestate}
	\rho(z)= \frac{1}{\pi^4 \sqrt{\mbox{Det}\Sigma}}e^{-z^T\Sigma^{-1}z}.
\end{equation}
Through a symplectic transformation on $Sp(8,\mathbb{R})=Sp(4,\mathbb{R})\otimes Sp(4,\mathbb{R})$., the $8\times 8$ covariance matrix $\Sigma$ corresponding to the bipartite state ~\eqref{sharestate}, can be written as 

\begin{eqnarray}
	\Sigma = \left(\begin{array}{cccccccc}
g_a & 0 & 0 & 0 & m_a & 0 & q_a & 0 \\
0 & g_a & 0 & 0 & 0 & m_c & 0 & q_b \\
0 & 0& g_b & 0 & q_a & 0 & m_b & 0 \\
0 & 0& 0 & g_b & 0 & q_b & 0 & m_d \\
m_a & 0 & q_a & 0 & g_c & 0 & 0 & 0 \\
0 & m_c & 0 & q_b & 0 & g_c & 0 & 0\\
q_a & 0 & m_b & 0 & 0 & 0 & g_d & 0 \\
0 & q_b & 0 & m_d & 0 & 0 & 0 & g_d 
	\end{array}\right),
\end{eqnarray}
For instance, for a pure state of two-mode symmetric Gaussian states for a bipartite system, the Covariance matrix can be reduced as
\begin{eqnarray}\label{covpure}
	\Sigma=\frac{b}{2}\left( \begin{array}{cc}
\mathbb{I}_4 & \gamma^T \\
\gamma & \mathbb{I}_4
	\end{array}\right),\;
\mbox{with}\; 
\gamma = \left(\begin{array}{cc}
n\mathbb{I}_2 & m\sigma_z \\
m\sigma_z & -n\mathbb{I}_2
\end{array}\right); \; m,n\in \mathbb{R},\; b>0.
\end{eqnarray}
  The characteristic polynomial $P_\Sigma(\lambda)=\mbox{Det}(\Sigma-\lambda\mathbb{I})$ of $\Sigma$
 has the roots
\begin{equation}
	\lambda^\Sigma_{\pm}= (2/b)(1\pm R), \mbox{with}\; R=\sqrt{m^2+n^2}\ge 0.
\end{equation}
Since $\Sigma$ is a covariance matrix, by definition it is positive definite, i.e., $\lambda^\Sigma_{\pm}\ge 0$. On the other hand, $b>0$ implies the only allowed value of $R$ is $0<R<1$. Hence,  the points $(m,n)$  lies inside the unit circle ($m^2+n^2<R^2$).
From now on, for simplicity, we assume
\begin{equation}
	b=(1+R)/(1-R),\; \mbox{with}\; R=\sqrt{m^2+n^2}.
\end{equation}
Now, one can utilize  the definition ~\eqref{gmunudefnnc} along with ~\eqref{gmunub},~\eqref{gmunu0} and ~\eqref{bmunu0} for the covariance matrix ~\eqref{covpure}, to compute the Fisher-Rao metric, which reads
\begin{eqnarray}
	g^0_{11}&=&\frac{4(1-m^2+n^2)}{(1-R^2)^2},\;\; b_{11}= \frac{16n^2}{R^2(R^2-1)(R+1)},\\
		g^0_{22}&=&\frac{4(1+m^2-n^2)}{(1-R^2)^2}, \;\; b_{22}= \frac{16m^2}{R^2(R^2-1)(R+1)},\\
		g^0_{12}&=& g^0_{21}= \frac{8mn}{(1-R^2)^2}, \;\; b_{12}= b_{21}=\frac{16mn}{R^2(R^2-1)(R+1)}.
\end{eqnarray}
Hence, we have the determinant $\Delta_g$ of the metric $[g_{\mu\nu}]=[g^0_{\mu\nu}+b_{\mu\nu}]$ as
\begin{equation}
	\Delta_g= \frac{16}{(1-R^2)^3}[1+(2-R)^2]
\end{equation}
Now if we wish to integrate $\Delta_g$ to have the volume of the possible quantum states, we see that $\int_{m^2+n^2<R^2} \Delta_g dmdn$ does not converge (as expected). Therefore, we need a regularization factor, which is obtained from ~\eqref{regularizationfunction}. In particular,
\begin{eqnarray}
	\Upsilon(\Sigma) &=& e^{-\frac{1}{\kappa}\mbox{Tr}[adj(\Sigma)]} \log[1+(\mbox{det}\Sigma)^m]
	=e^{-\tau/\kappa} \log[1+b^2\tau^2 (1-R^2)^2/2^8],\\
&&	\mbox{with}\; \tau=\mbox{Tr}[\mbox{Adj}\Sigma]=(1-R^2)^3 b^7/16.
\end{eqnarray} 
Note that, the volume measure contains an arbitrary scale factor $\kappa$.  Since we are interested only in the relative measures of volume, we choose it suitably. For instance,  for $\kappa=2$, we have 
$	\int_{m^2+n^2\le 1} \Upsilon(\Sigma)\Delta_g dm dn \approx 1.95268$. 
\begin{figure}
	\includegraphics[width=8cm]{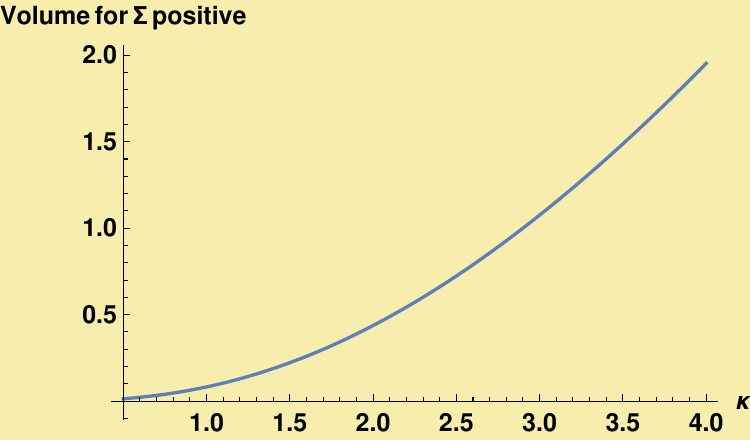}
	\caption{Measure of  states for which $\Sigma>0$, with respect to scale factor $\kappa$. Here we vary $\kappa$ from $1/2$ to $4$. }
	\label{vol_sigpositive}
\end{figure}
Figure-FIG.~\ref{vol_sigpositive} represents the variation of allowed volume for positive $\Sigma$, with respect to the scale factor $\kappa$.\\
However, for a bonafide covariance matrix in quantum mechanics, $\Sigma$ has to satisfy RSUP ~\eqref{NCRSUP}, which can be restated as that the smallest eigenvalue of $2i\Omega^{-1}\Sigma$ should be greater than one ($\nu_-\ge 1$). On the other hand, for separable states,  the smallest eigenvalue $\nu_-'$ of $2i(\Omega')^{-1}\Sigma'$ (after mirror reflection/partial transpose) should also be greater than one ($\nu_-'\ge 1$). Explicitly written, $\nu_{-}$ and $\nu_{-}'$ are given by
\begin{eqnarray}
	\nu_{-} &=& \frac{b}{(1-\eta\theta)\sqrt{2}} \sqrt{\omega_{-} - \sqrt{\omega_-^2 - 4(1-\eta\theta)^2(1-R^2)^2}}, \label{numinus}\\
	\nu_{-}' &=& \frac{b}{(1-\eta\theta)\sqrt{2}} \sqrt{\omega_{+} - \sqrt{\omega_+^2 - 4(1-\eta\theta)^2(1-R^2)^2}},\label{nuprimeminus},
	\end{eqnarray}
where
\begin{eqnarray} 
	\omega_{\pm} =2(1\pm\eta^2)+ (1\mp n^2) (\eta^2 +\theta^2) \pm 2 (1+\eta\theta) m^2 \nonumber \\
 	+ n (1\mp 1) (\eta^2 -\theta^2)  + 2m (1\pm 1) (\eta+\theta). 
\end{eqnarray}
Note that for $\theta=\eta=0$, the equations ~\eqref{numinus} and ~\eqref{nuprimeminus} are reduced to $\nu_-\vert_{\theta,\eta\to 0}=b\sqrt{1-R^2}\ge 1$ and $ \nu_-'\vert_{\theta,\eta\to 0}=1+R >1$, respectively. That means the states are always separable in commutative space. We choose our covariance matrix in such form that the form of covariance matrix is separable in commutative space so that we can focus on the entanglement induced solely by NC-parameters. \\
Now let us summarize the parameter space for bonafide quantum states, separable and entangled states as follows.
\begin{equation}\label{thetaquantumexample}
	\Theta_{\mbox{quantum}}^{NC}:= \left\{\theta =(m,n) \in \mathbb{R}^2 \vert m^2+n^2<1,  \nu_{-}\ge 1 \right\},
\end{equation}
whereas for separable state 
\begin{equation}\label{thetaseparableexample}
	\Theta_{\mbox{separable}}^{NC}:= \left\{\theta =(m,n) \in \mathbb{R}^2 \vert m^2+n^2<1,  \nu_{-}'\ge 1\right\}.
\end{equation}
If we assume that, a quantum state is entangled if it is not separable, then
\begin{equation}\label{thetaentangledexample}
	\Theta_{\mbox{entangled}}^{NC}:= \Theta_{\mbox{quantum}}^{NC}-\Theta_{\mbox{separable}}^{NC}.
\end{equation}  
Now the volume $\Gamma_{\mbox{quantum}}^{\mbox{NC}}$ and $\Gamma_{\mbox{separable}}^{\mbox{NC}}$ of total quantum state and separable state, respectively in NC-space are given by
\begin{eqnarray}
	\Gamma_{\mbox{quantum}}^{\mbox{NC}}= \int_{(m,n)\in \Theta_{\mbox{quantum}}^{NC}}  \sqrt{\Delta_g}\Upsilon(\Sigma) dmdn,\\
		\Gamma_{\mbox{separable}}^{\mbox{NC}}= \int_{(m,n)\in \Theta_{\mbox{separable}}^{NC}}  \sqrt{\Delta_g}\Upsilon(\Sigma) dmdn.
\end{eqnarray}
Hence, the volume $\Gamma_{\mbox{entangled}}^{\mbox{NC}}$  of total entangled quantum state is 
\begin{equation}
	\Gamma_{\mbox{entangled}}^{\mbox{NC}}= \Gamma_{\mbox{quantum}}^{\mbox{NC}} - 	\Gamma_{\mbox{separable}}^{\mbox{NC}}.
\end{equation}
To get a glimpse of how the volume measure of states varies with NC parameters, we plot it in figure- FIG.~\ref{volrelative} for momentum NC parameter $\eta\to 0$, i.e., for the case of only spatial noncommutativity ($\theta \neq 0$). We see that, the number of entangled states increases with the NC-parameter $\theta$, as expected.
\begin{figure}
	\includegraphics[width=8cm]{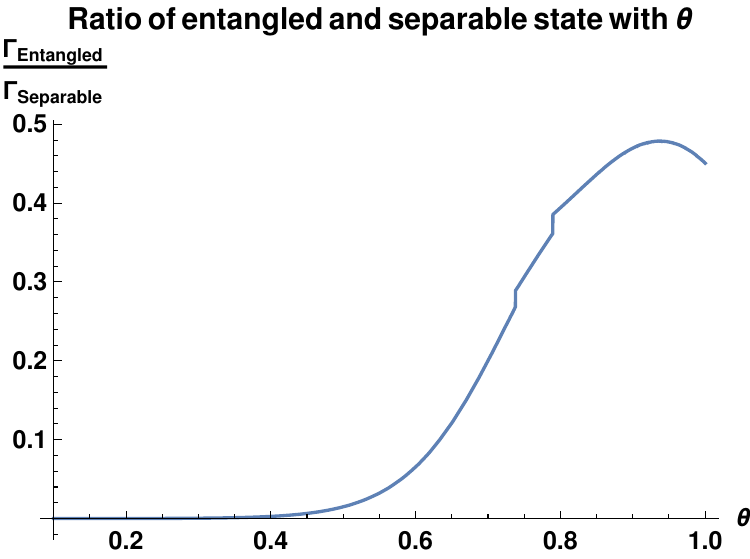}
	\caption{\textcolor{blue}{The ratio of the measure of entangled states and separable states with respect to NC parameters $\theta $. Here we take scale factor $\kappa \to 4$. We vary $\theta$ from $0.1$ to $1$. The measure of the entangled state increases with the NC parameter since the entanglement is due to deformation in phase-space.}}
	\label{volratiotheta}
\end{figure}
\begin{figure}
		\includegraphics[width=8cm]{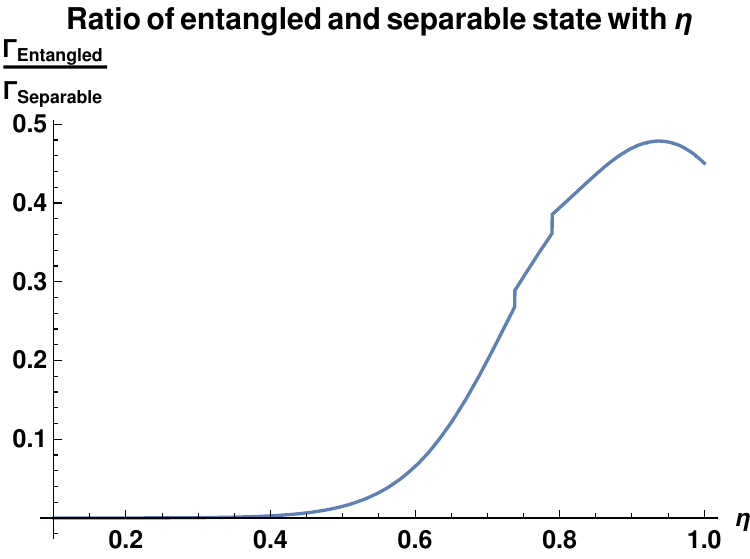}
\caption{\textcolor{blue}{The ratio of the measure of entangled states and separable states with respect to NC parameter  $\eta$. Here we take scale factor $\kappa \to 4$. We vary  $\eta$from $0.1$ to $1$. The measure of the entangled state increases with the NC parameter, as expected.} }
	\label{volratioeta}
\end{figure}
In figure-FIG~\ref{volratiotheta} and FIG~\ref{volratioeta}, we plot the ratio of the volume of entangled states and separable states, with respect to NC parameters $\theta$ and $\eta$, respectively. One can see that it varies in the same manner with respect to $\theta$ and $\eta$. 
%%%%%%%%%%%%%%%%%%%%%%%%%%%%%%%%%%%%%%%%%%%%%%%%%%%%%%%
%%%%%%%%%%%%%%%%%%%%%%%%%%%%%%%%%%%%%%%%%%%%%%%%%%%%%%
\section{Conclusions}
In the present paper, we tackled the problem of evaluating the effects of noncommutative (NC) space parameters on the volume of bipartite Gaussian states. With the help of formalisms of information geometry, we have constructed all three associated manifolds, one for all possible quantum states, another for separable states, and one for entangled states. For toy models, we have considered the bipartite Gaussian state, whose commutative space counterpart is separable, so that we can envisage how   NC-parameters $\theta,\eta$ affect the entanglement property. Since, NC-quantum mechanics can be mapped onto an equivalent problem of Landau level, where the external magnetic field plays a role as the momentum-momentum NC-parameter, the present piece of theoretical work opens up an avenue for experimental verification through a lower-dimensional condensed matter system. Moreover, since the entanglement of Gaussian states has been shown to be created both via position-position and momentum-momentum noncommutativity, the one-particle sector of theories can exhibit the experimental effects shown in the present work.
\section{Acknowledgement}
S. Nandi is grateful to ANRF (Formerly SERB), Govt. of India, for fellowship support through project grant no  EEQ/2023/000784. P. Patra is grateful to ANRF (Formerly SERB), Govt. of India, for financial support through project grant No. EEQ/2023/000784.

\end{document}